\documentclass[aps,prd,10pt,a4paper,superscriptaddress,preprintnumbers,showpacs,notitlepage,eqsecnum,nofootinbib]{revtex4-1}
\usepackage[english]{babel}
\usepackage{graphicx}
\usepackage{amsmath}
\usepackage{color}
\usepackage{amsmath}
\usepackage{amsfonts}
\usepackage{verbatim}
\usepackage{amssymb}
\usepackage{graphicx}
\usepackage{epstopdf}
\usepackage{mathrsfs}%

\def\sqr#1#2{{\vcenter{\vbox{\hrule height.#2pt\hbox{\vrule
width.#2pt height#1pt \kern#1pt\vrule width.#2pt}\hrule height.#2pt}}}}

\begin{document}

\title{On the mechanical stability of asymptotically flat black holes\\
with minimally coupled scalar hair\\
}
\author{Andr\'{e}s Anabal\'{o}n}
\affiliation{Departamento de Ciencias, Facultad de Artes Liberales y
Facultad de Ingenier\'{i}a y Ciencias, Universidad Adolfo Ib\'{a}\~{n}ez, Vi\~{n}a del Mar, Chile.
}

\author{Nathalie Deruelle}

\affiliation{
APC, CNRS-Universit\'e 
Paris 7, 75205 Paris CEDEX 13, France.}

\begin{abstract}
We show that the asymptotically flat hairy black holes, solutions of the
Einstein field equations minimally coupled to a scalar field, previously
discovered by one of us, present mode instability against linear radial
perturbations. It is also shown that the number of unstable modes is finite
and their frequencies can be made arbitrarily small.
\end{abstract}

\maketitle

\section{Introduction}

For any number of minimally coupled scalar fields, the most general form of
the no hair theorem, in asymptotically flat spacetimes, states that when the
scalar field potential is everywhere non-negative then there are no
non-trivial regular black hole spacetimes (see \cite{Sudarsky:1995zg,
Nucamendi:1995ex} and references therein). Therefore, to have an
asymptotically flat hairy black hole with a minimally coupled scalar field a
necessary condition is a scalar field potential with a negative region. In 
\cite{Andrespapers1, Andrespapers2, Andrespapers3}  a family of static, spherically symmetric solution of the
Einstein field equations with a minimally coupled scalar field $%
\phi $ with a non-trivial potential $V(\phi )$ was found and analyzed. For a whole range of the
parameters this solution describes either asymptotically flat or (anti) de
Sitter black holes. The main subject of the present paper is to analyze the mode
stability of the asymptotically flat black holes, whose properties are
summarized in section I below. The scalar field potential in the
asymptotically flat case is depicted in figure~\ref{fig:PotAndres}.

The question of mode stability is already non-trivial for vacuum solutions,
which  was settled in the static case by Regge-Wheeler and Zerilli (for
references and an account of these results see \cite{Chandrasekhar:1985kt}).
It was shown in these seminal works  that the stability problem can be
mapped to the analysis of the spectrum of a Schr\"{o}dinger operator. An
everywhere positive spectrum implies there are no modes which exponentially grow in time. Since, as can be seen from figure~\ref%
{fig:PotAndres}, the potential $V(\phi )$ is not bounded from below one
would thus expect  that something should go very
wrong with the stability of the solution.
This is the issue we address in this paper. Using the method of Bronnikov et 
\textsl{al.} \cite{Bronnikov1, Bronnikov2} we shall study the equations of
motion for the radial perturbations in section \ref{sec:stability}. We will
see that the effective potential $V_{\mathrm{eff}}(\rho )$ in which these
modes $\delta \phi (t,\rho )$ propagate, 
\begin{equation}
-{\frac{d^{2}u}{d\rho ^{2}}}+V_{\mathrm{eff}}u=E^{2}u\qquad \hbox{with}%
\qquad \delta \phi \propto e^{\mathrm{i}Et}u(\rho )\,,  \label{schroe}
\end{equation}%
where $\rho $ is a \textquotedblleft tortoise" radial coordinate sending the
horizon at minus infinity, always exhibits a negative region. A sufficient
condition for the existence of bound states with negative $E^{2}$ (for
bounded $V_{\mathrm{eff}}$ that fall-off faster than $\left\vert \rho
\right\vert ^{-2}$)\ is the Simon criteria \cite{Simon1, Simon2}, which
states that whenever the integral

\begin{equation}
S\equiv \int_{-\infty }^{+\infty }V_{\mathrm{eff}}\,d\rho
\end{equation}%
is negative there will always be at least one bound state with negative $%
E^{2}$. Therefore, we only study effective potentials with a positive Simon
integral. Using standard \textquotedblleft shooting" techniques to solve (%
\ref{schroe}), we do indeed find unstable
modes.

What is remarkable however is that this instability is somehow marginal.
Indeed, as we shall comment upon in section~\ref{sec:savethephenomena},
there is only a finite number of unstable modes and, moreover, as follows
from the sharp Lieb-Thirring inequality in one dimension \cite%
{Hundertmark:1998cz}, their characteristic time of growth can be made
arbitrarily large for certain values of the black holes parameters, as is
the case if the size of the black hole is small enough. 

The metric signature is $(-,+,+,+)$ and we  set $c=1=\kappa $. This
implies that a canonically normalized scalar field, $\phi ,$ is
dimensionless.

\section{\label{sec:BHsol}Hairy black hole solutions}

\begin{figure}[tbp]
\includegraphics[scale=0.7]{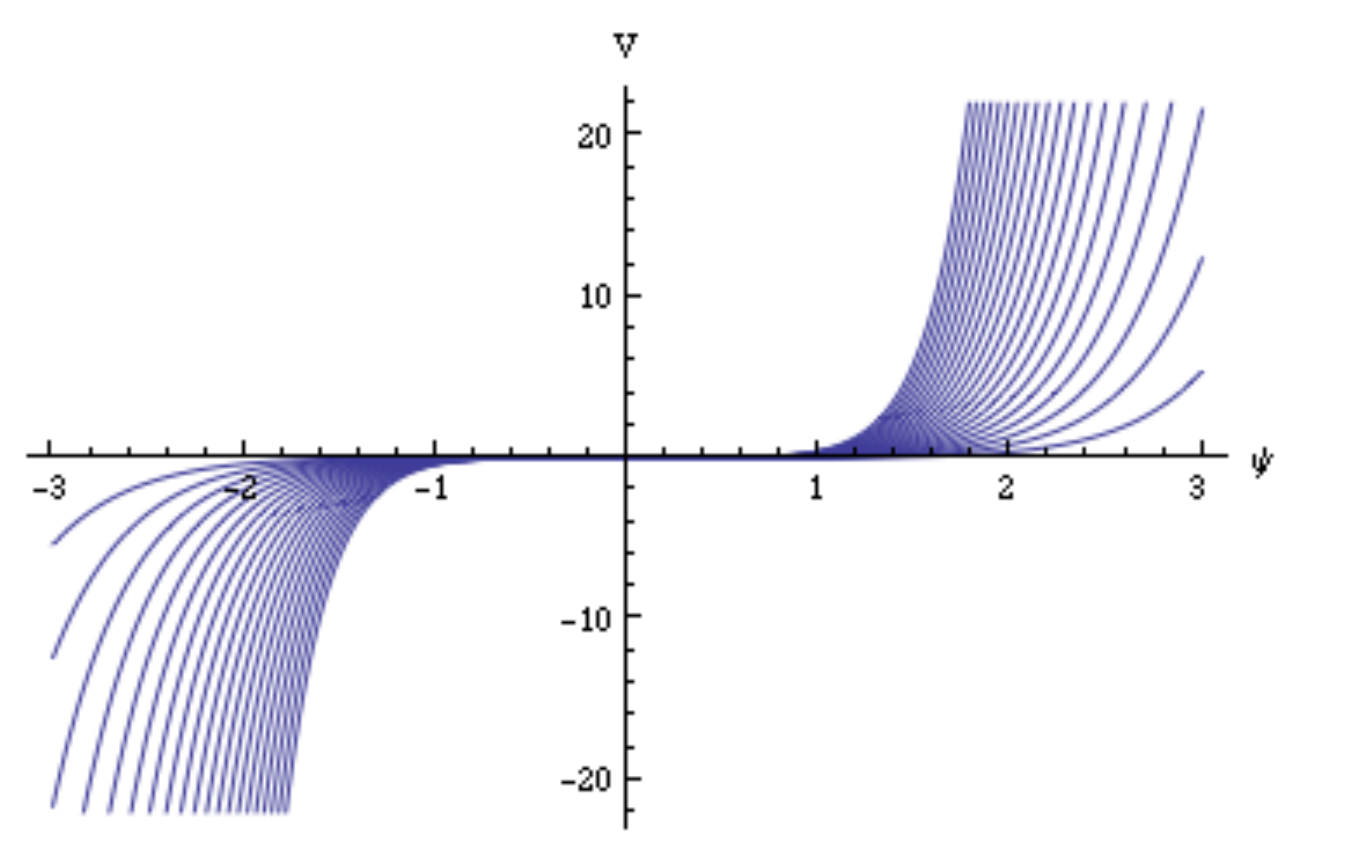}
\caption{The potential (2.3) for various values of the parameter $\protect\nu%
>1$ (and $\protect\alpha>0$). }
\label{fig:PotAndres}
\end{figure}

Here the main results obtained in \cite{Andrespapers1, Andrespapers2,
Andrespapers3} are summarized. Consider the Einstein field equations minimally coupled to a scalar field 
\begin{equation}
G_{\mu \nu }=\partial _{\mu }\phi \,\partial _{\nu }\phi -g_{\mu \nu }\left( 
{\frac{1}{2}}g^{\rho \sigma }\partial _{\rho }\phi \,\partial _{\sigma }\phi
+V(\phi )\right)  \label{EFEQ}
\end{equation}%
which through the Bianchi identity imply the
Klein-Gordon equation, 
\begin{equation}
\mathchoice{\vcenter{\vbox{\hrule height.4pt\hbox{\vrule
width.4pt height 5pt \kern 5pt\vrule width.4pt}\hrule height.4pt}}}{\vcenter{\vbox{\hrule height.4pt\hbox{\vrule
width.4pt height 5pt \kern 5pt\vrule width.4pt}\hrule height.4pt}}}{\vcenter{\vbox{\hrule height.3pt\hbox{\vrule
width.3pt height 2.1pt \kern 2.1pt\vrule width.3pt}\hrule height.3pt}}}{\vcenter{\vbox{\hrule height.3pt\hbox{\vrule
width.3pt height 1.5pt \kern 1.5pt\vrule width.3pt}\hrule height.3pt}}}\phi
-V_{\phi }=0\,.  \label{KGEQ}
\end{equation}%
where $V_{\phi }\equiv dV/d\phi $. It was shown in \cite{Andrespapers1} that
the most general\footnote{%
Excluding a cosmological constant.} potential compatible with the
Carter-Debever-Pleba\'{n}ski ansatz is

\begin{equation}
V(\psi)={\frac{\alpha}{\nu^2}}\left\{{\frac{\nu-1}{\nu+2}}\sinh[(1+\nu)\psi]+%
{\frac{\nu+1}{2-\nu}}\sinh[(\nu-1)\psi]-4{\frac{\nu^2-1}{4-\nu^2}}%
\sinh\psi\right\}\quad\hbox{with}\quad \psi\equiv\sqrt{\frac{2}{\nu^2-1}}%
\,\phi\,,
\end{equation}
where $\alpha$, which has dimension $length^{-2}$, and $\nu>1$ are the
parameters of the model. It is quite flat at the origin ($%
V(\psi)\simeq\alpha(\nu^2-1)\psi^5/30$) and is unbounded from below, see
figure~\ref{fig:PotAndres}.\newline

The equations (1-3) possess the following 4-dimensional, static and
spherically symmetric solutions 
\begin{equation}
\begin{aligned} \psi&=\ln x\qquad,\qquad
ds^2=\Omega(x)\left[-F(x)dt^2+{\eta^2 \,dx^2\over
F(x)}+d\theta^2+\sin^2\theta d\phi^2\right]\\ \hbox{with}\quad
\Omega(x)&={\nu^2x^{\nu-1}\over\eta^2(1-x^\nu)^2}\quad\hbox{and}\quad
F(x)=\eta^2\,{x^{2-\nu}(1-x^\nu)^2\over\nu^2}-\alpha\left[{1\over4-%
\nu^2}+{x^2\over\nu^2}\left(1-{x^{-\nu}\over2-\nu}-{x^\nu\over2+\nu}\right)\right]%
\,, \end{aligned}  \label{bh}
\end{equation}%
where $\eta >0$ is the unique integration constant of the solution. This
solution coincides with the Schwarzschild solution when $\nu =1$ (for
details see \cite{Andrespapers2}). The coordinate $x$ is dimensionless,
related to the standard Droste radial coordinate $r$  as 
\begin{equation}
r={\frac{\nu }{\eta }}\,{\frac{x^{\frac{\nu -1}{2}}}{|1-x^{\nu }|}}\,.
\label{droste}
\end{equation}%
Spatial infinity is at $x=1$. The solutions have two branches: $0<x<1$, with
the curvature singularity at $x=0$; and $x>1$, with the curvature
singularity at $x=\infty $. Since $\Omega F\rightarrow 1$, $\eta ^{2}\Omega
/F(dx/dr)^{2}\rightarrow 1$ and $\Omega \approx r^{2}$ when $x\rightarrow 1$%
, the metric is asymptotically flat; and actually it is asymptotically
Schwarzschild's metric with PPN parameters $\gamma _{\mathrm{PPN}}=\beta _{%
\mathrm{PPN}}=1$. \newline

The gravitational mass (read off from the asymptotic behaviour of the $%
g_{tt}=-(1-2m/r+\cdots )$ component of the metric) and the inertial mass of
the solutions (as defined for example by the Komar integral or the Katz
superpotential, see \cite{Andrespapers2} and \cite{KomarKatz}) are equal,
and given by 
\begin{equation}
m=\pm {\frac{\alpha +3\eta ^{2}}{6\eta ^{3}}}  \label{mass}
\end{equation}%
with the upper sign for the branch $x<1$ and the lower for the branch $x>1$.
The mass is positive for all $\alpha >-3\eta ^{2}$ for the $x<1$ branch and
for all $\alpha <-3\eta ^{2}$ for the branch $x>1$. \newline

The metric (\ref{bh}) represents a Schwarzschild-type black hole if the
metric function $F(x)$ has one single zero, if it is positive at spatial
infinity, and negative at the curvature singularity. These conditions are
fulfilled

(a) when $\alpha >0$: for all $\eta $, if $\nu \in \lbrack 1,2]$; and for $%
\eta ^{2}-\frac{\alpha }{\nu -2}<0$ if $\nu >2$. In those cases the horizon $%
x_{+}$ is such that $0<x_{+}<1$ and the mass of the black hole is (2.6) with
a plus sign;

(b) when $\alpha <0$: if $\eta ^{2}+\frac{\alpha }{\nu +2}<0$. Then the
horizon $x_{+}>1$ and the mass is (\ref{bh}) with a minus sign.

By definition, the horizon $x_{+}$ of the black hole is such that $%
F(x_{+})=0 $ with $F$ given in (1.4)~:  
\begin{equation}
\eta ^{2}=\alpha {\frac{\nu ^{2}x_{+}^{\nu }-(\nu +2)x_{+}^{2}+(4-\nu
^{2})x_{+}^{2+\nu }-(2-\nu )x_{+}^{2+2\nu }}{(4-\nu
^{2})x_{+}^{2}(1-x_{+}^{\nu })^{2}}}\,.  \label{eta}
\end{equation}%
The asymptotic behaviours of $\eta^2$ are~:

\begin{equation}
 \left\{\begin{aligned}
{\eta^2\over\alpha}&\approx {1-x_+\over3}\quad\hbox{for}\quad x_+\to1\\
{\eta^2\over\alpha}&\approx-{1\over2+\nu}(1-\nu x_+^{-\nu})\ \hbox{for}\  x_+\to\infty
 \end{aligned}\right.
 \quad,\quad 
 \left\{\begin{aligned}
{\eta^2\over\alpha}&\approx{\nu^2\,x_+^{\nu-2}\over4-\nu^2}\to+\infty\quad\hbox{if}\quad \nu\in[1,2]\\
{\eta^2\over\alpha}&\approx{1\over\nu-2}\left(1-{\nu^2 x_+^{\nu-2}\over\nu+2}\right)\quad\hbox{if}\quad \nu>2
 \end{aligned}\right.
 \quad\hbox{for}\quad x_+\to0\,.
 \label{eta1}
 \end{equation}

 Besides these values of $x_{+}$, there is a further interesting value given
by the condition $\frac{d(\eta) ^{2}}{dx_{+}}=0$, whose numerator vanishes if%
\begin{equation}
\frac{\nu -2}{\nu +2}=\frac{x_{+}^{\nu }-x_{+}^{2}}{x_{+}^{\nu +2}-1}.
\label{eq1}
\end{equation}%
Note that $x_{+}=1$ is a solution of (\ref{eq1}), however it is not a
critical point since $\left. \frac{d\eta ^{2}}{dx_{+}}\right\vert
_{x_{+}=1}=-\frac{\alpha }{3}$. Actually, since the LHS of (\ref{eq1}) is
such that $-\frac{1}{3}<\frac{\nu -2}{\nu +2}<1$ and the numerator as well
as the denominator of the RHS of (\ref{eq1}) are monotonically increasing or
decreasing it follows that (\ref{eq1}) can have at most one solution, which
is $x_{+}=1$. Therefore, there are no critical points of the function (\ref%
{eta}) in the relevant interval and it is possible to parameterize
the solution in terms of $x_{+}$ instead of $\eta $. The conditions (a,b)
for the existence of an horizon can thus be replaced by (a'): $\alpha >0$, $%
x_{+}\in (0,1)$; (b'): $\alpha <0$, $x_{+}>1$.

The mass (\ref{mass}) of the black hole then becomes a function of $x_{+}$
or alternatively $r_{+}$ through (\ref{droste}) and (\ref{eta}). Then, it is
possible to show that close to spacelike infinity we recover the
Schwarzschild relation 
\begin{equation}
\lim_{x_{+}\rightarrow 1}\frac{2m}{r_{+}}=1
\end{equation}%
and close to the singularities we have, using (\ref{eta1}) 
\begin{equation}
\left\{ \begin{aligned}& \lim_{x_{+}\rightarrow 0}\,mr_+^{\nu-2}=
(4-\nu^2)^{(\nu-1)/2}(\sqrt\alpha)^{1-\nu}/(2\nu)\qquad\hbox{for}\quad%
\alpha>0\quad\hbox{and}\quad \nu\in[1,2]\,,\\ &\lim_{x_{+}\rightarrow
0}\,m=(\nu+1)\sqrt{\nu-2}/(6\sqrt{\alpha})\qquad\hbox{for}\quad\alpha>0%
\quad\hbox{and}\quad \nu>2\,,\\ &\lim_{x_{+}\rightarrow
\infty}\,m=\to(\nu-1)\sqrt{\nu+2}/(6\sqrt{-\alpha})\qquad\hbox{for}\quad%
\alpha<0 \,.\end{aligned}\right.
\end{equation}%
We note that the mass can tend to a constant when the size of the black hole
shrinks to zero.

\section{\label{sec:stability}Linear instability}

\subsection{The Bronnikov \textsl{et al.} equations for linear radial
perturbations}

We summarize here the results obtained in \cite{Bronnikov1, Bronnikov2}.
Consider the metric 
\begin{equation}
ds^{2}=-e^{2[\gamma _{0}(x)+\delta \gamma (t,x)]}dt^{2}+e^{2[\alpha
_{0}(x)+\delta \alpha (t,x)]}dx^{2}+e^{2\lambda _{0}(x)}(d\theta ^{2}+\sin
^{2}\theta \,d\phi ^{2})
\end{equation}%
and scalar field $\phi =\phi _{0}(x)+\delta \phi (t,x)$. Expand the Einstein
and Klein-Gordon equations (\ref{EFEQ}) and (\ref{KGEQ}) to linear order in
the perturbations $\delta \gamma $, $\delta \alpha $ and $\delta \phi $.
After rearrangement and proper use of the zeroth order, background,
equations, they boil down to three independent equations. Two of them are
constraints.
\begin{equation}
\begin{aligned}
\delta\alpha&={\phi_0'\over2\lambda_0'}\,\delta\phi\\
\delta\gamma'&={1\over2\lambda_0'^2}\left[(\phi_0'e^{-2\lambda_0}-\phi_0'V-\lambda_0'V_{\phi})e^{2\alpha_0}\delta\phi+\lambda_0'\phi_0'\,\delta\phi'\right]
\end{aligned}
\end{equation}
where a prime denotes derivation with respect to $x$ and where $V$ and $V_\phi$ are evaluated on the background solution $\phi_0$.

 The third is an equation of propagation for $\delta \phi $,
which, after setting 
\begin{equation}
\delta \phi \equiv e^{\mathrm{i}Et}e^{-\lambda _{0}}u(x)\,,  \label{FACT}
\end{equation}%
can be put under the canonical form 
\begin{equation}
{\frac{d^{2}u}{d\rho ^{2}}}+(E^{2}-V_{\mathrm{eff}})u=0  \label{RP}
\end{equation}%
where $E$ is the \textquotedblleft energy" of the mode, where $\rho $ is the
\textquotedblleft tortoise coordinate" such that 
\begin{equation}
\rho ^{\prime }\equiv {\frac{d\rho }{dx}}=e^{\alpha _{0}-\gamma _{0}}
\end{equation}%
and where the effective potential is given by 
\begin{equation}
V_{\mathrm{eff}}={\frac{e^{2\gamma _{0}}}{\lambda _{0}^{\prime 2}}}\left[
2\lambda _{0}^{\prime }\phi _{0}^{\prime }V_{\phi }-e_{0}^{-2\alpha
_{0}}\lambda _{0}^{\prime 4}+\phi _{0}^{\prime 2}(V-e^{-2\lambda
_{0}})+\lambda _{0}^{\prime 2}(e^{-2\lambda _{0}}-V+V_{\phi \phi })\right]
\,.
\end{equation}

\subsection{The effective potential}

The scalar field potential $V$ is given in (2.3) and the background solution
is 
\begin{equation}
\gamma _{0}=\ln \sqrt{\Omega F}\quad ,\quad \alpha _{0}=\ln \eta \sqrt{\frac{%
\Omega }{F}}\quad ,\quad \lambda _{0}=\ln \sqrt{\Omega },\quad ,\quad \phi
_{0}=\sqrt{\frac{\nu ^{2}-1}{2}}\psi
\end{equation}%
with $\Omega $, $F$ and $\psi $ given in (2.4). Inserting (3.6) into (3.5)
one finds 
\begin{equation}
V_{\mathrm{eff}}=F{\frac{\mathrm{N}_{1}+x^{2}\mathrm{N_{2}}}{\mathrm{D}}}
\label{EP}
\end{equation}%
and where the explicit expressions for $\mathrm{N}_{1}$, $\mathrm{N}_{2}$
and $\mathrm{D}$ are given in Appendix~\ref{app:Veff}. 

Let us now look at the shape of the effective potential $V_{\mathrm{eff}}$.
From its expression given in Appendix~\ref{app:Veff} we have that, near
spatial infinity is simply 
\begin{equation}
V_{\mathrm{eff}}=\left( \alpha -\eta ^{2}\right) \left( x-1\right)
^{3}+O(x-1)^{4}=O(\rho ^{-3}).  \label{BC1}
\end{equation}%
Note that the condition (b') for the existence of the horizon implies that
whenever $x>1$ then $\alpha <0\Longrightarrow $ $V_{\mathrm{eff}}$ is
negative close to spacelike infinity. The condition (a'), $\alpha >0$ and $%
x<1$ and the use of (\ref{eta1}) shows that it is
asymptotically negative if $\nu >3$; for $\alpha >0$ and $\nu \in \lbrack
1,3]$ it can be positive close to spacelike infinity for small enough values
of $x_{+}$.

Around the black hole horizon 
\begin{equation}
d\rho =\frac{\eta dx}{F}=\frac{\eta dx}{(x-x_{+})F^{\prime }(x_{+})}%
+O(x-x_{+})^{-2},
\end{equation}%
therefore,%
\begin{equation}
x>1\Longrightarrow \lim_{x\rightarrow x_{+}}\rho =\lim_{x\rightarrow x_{+}}%
\frac{\eta }{F^{\prime }(x_{+})}\ln \left\vert x-x_{+}\right\vert =+\infty
,\qquad x<1\Longrightarrow \lim_{x\rightarrow x_{+}}\rho =\lim_{x\rightarrow
x_{+}}\frac{\eta }{F^{\prime }(x_{+})}\ln \left\vert x-x_{+}\right\vert
=-\infty
\end{equation}%
The effective potential (\ref{EP}) is overall multiplied by $F$, which
implies that around the horizon

\begin{equation}
V_{\mathrm{eff}}(x=x_{+})\sim \exp \left( -\left\vert F^{\prime }(x_{+})\eta
^{-1}\rho \right\vert \right)  \label{BC2}
\end{equation}%
From (\ref{BC1}) and (\ref{BC2}) it follows that the number of bound states
with negative $E^{2}$ should be finite, as is the case with any potential
tha fall-off faster than $\left\vert \rho \right\vert ^{-2}$ in one
dimension \cite{Reed:1979ne}.

Near the horizon we have 
\begin{equation}
V_{\mathrm{eff}}={\frac{dV_{\mathrm{eff}}}{dx}}\bigg\vert_{x_{+}}(x-x_{+})+%
\cdots \quad \hbox{with}\quad {\frac{dV_{\mathrm{eff}}}{dx}}\bigg\vert%
_{x_{+}}={\frac{dF}{dx}}{\frac{\mathrm{N_{1}}+x^{2}\mathrm{N_{2}}}{\mathrm{D}%
}}\bigg\vert_{x_{+}}
\end{equation}%
where $\mathrm{N_{1}}$, $\mathrm{N_{2}}$ and $D$ are given in Appendix~\ref%
{app:Veff}. The value of ${dV_{\mathrm{eff}}/dx}$ at $x=x_{+}$ is depicted
in figure 2. For $\alpha >0$ it is negative unless the
horizon size is small enough, all the more so as $\nu \rightarrow 1$. For $%
\alpha <0$ it is always negative for $\nu \in \lbrack 1,3]$, and can be
positive if $\nu >3 $ and the horizon size is big enough all the more so as $%
\nu $ increases. 
\begin{figure}[tbp]
\includegraphics[scale=0.4]{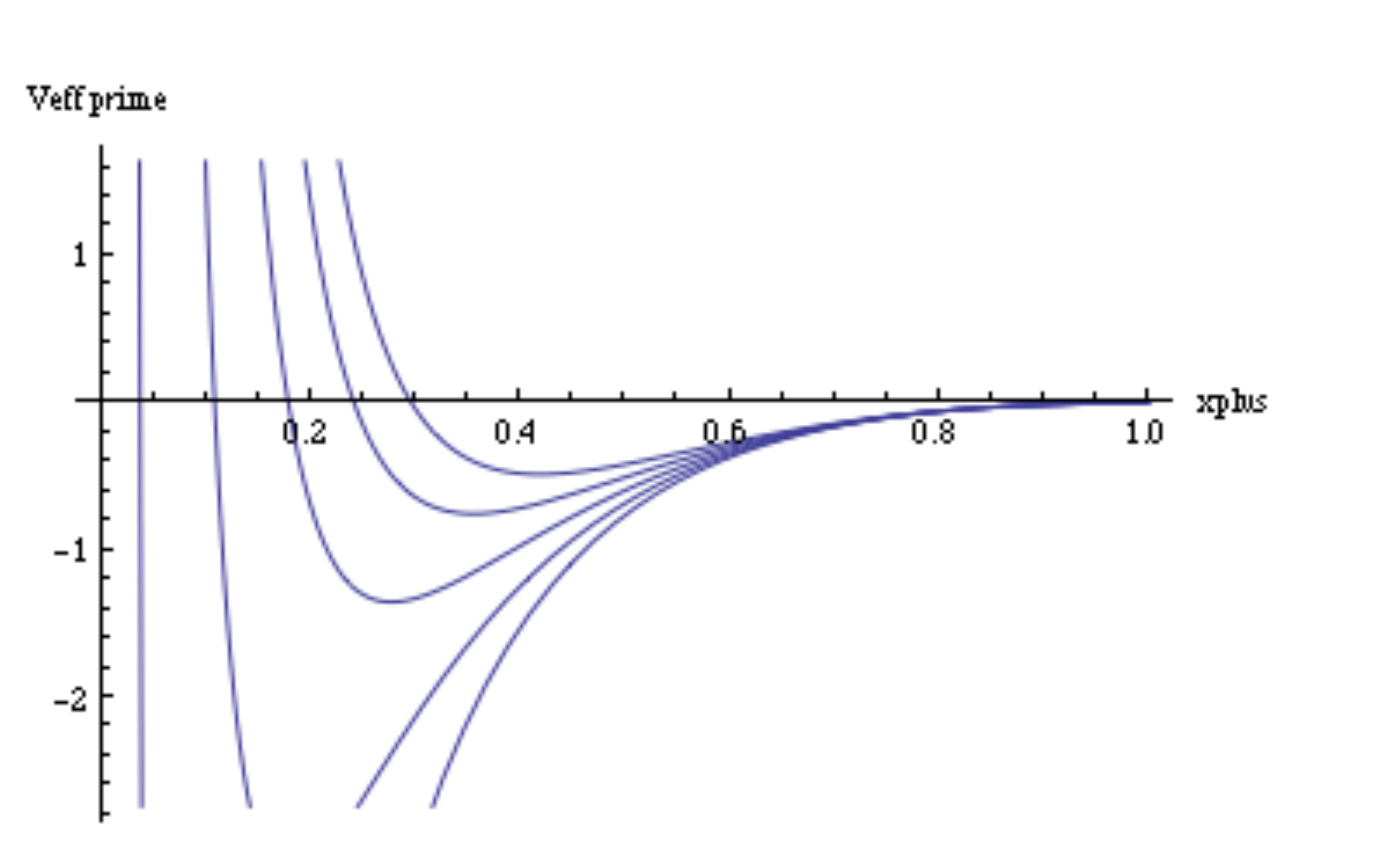} %
\includegraphics[scale=0.4]{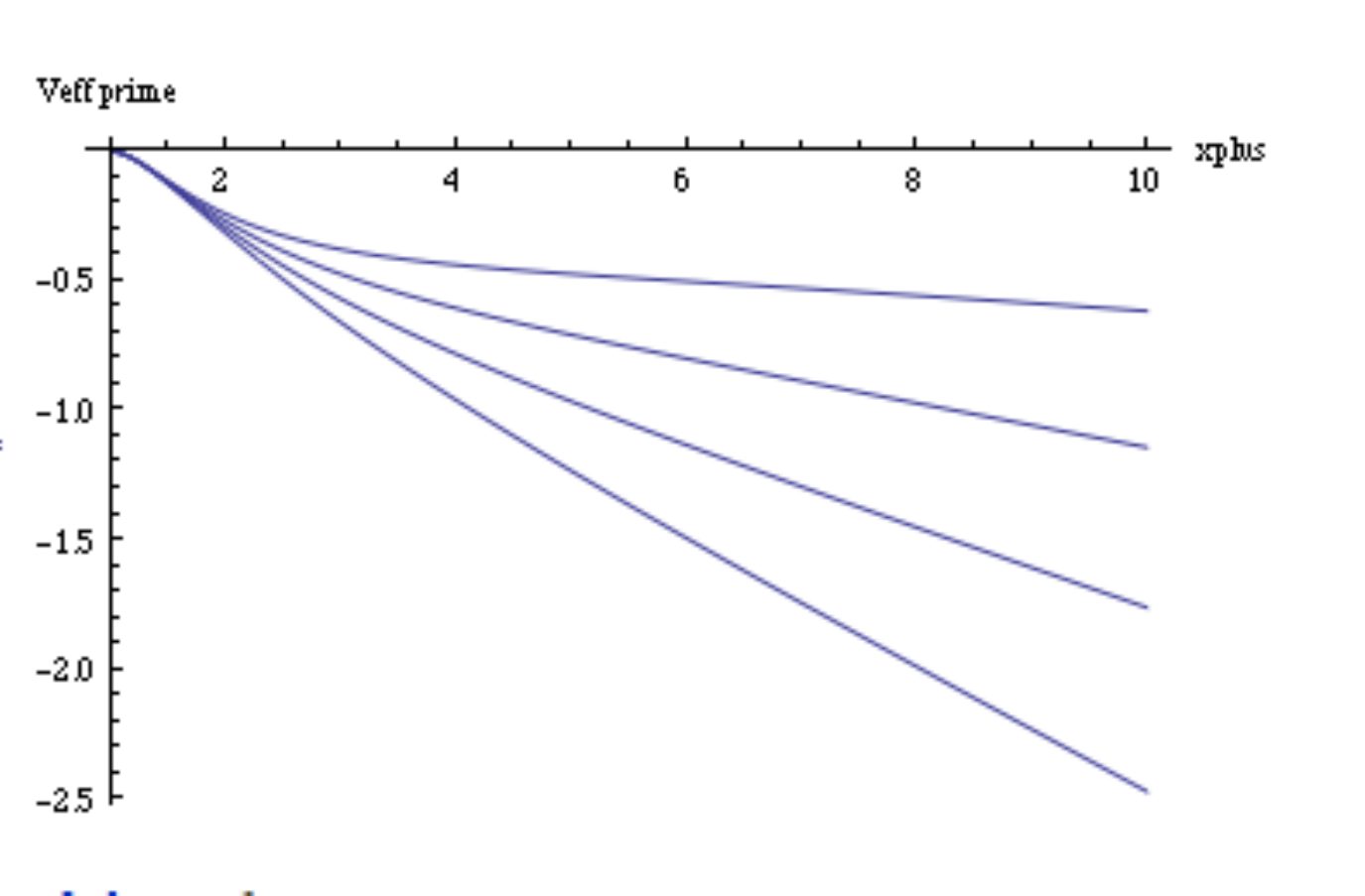} %
\includegraphics[scale=0.4]{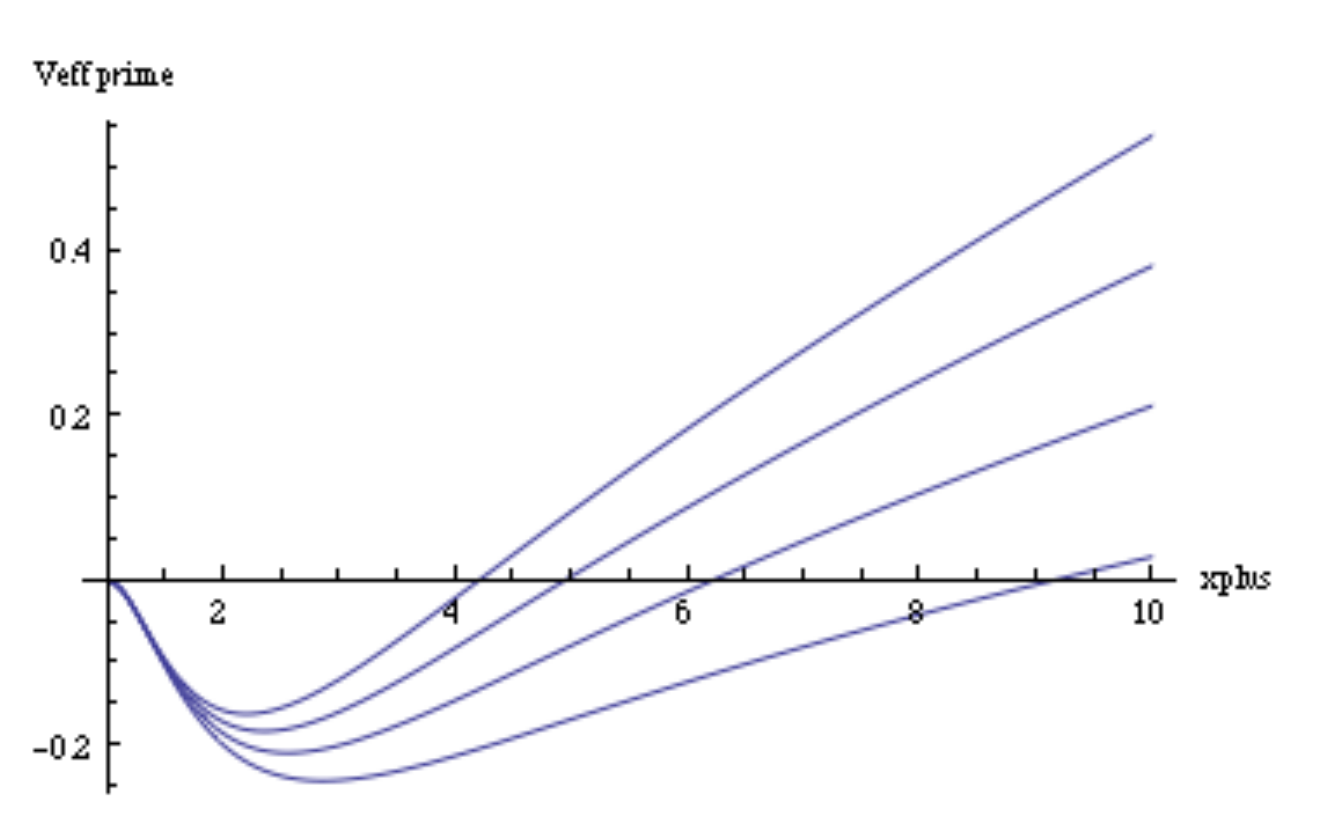}
\caption{${\frac{dV_{\mathrm{eff}}}{dx}}\big\vert_{x_{+}}$ as a function of
the horizon location $x_{+}$ for various values of $\protect\nu $. Left
panel~: $\protect\alpha >0$ and increasing values of $\protect\nu >1$ from
right to left. Middle panel~: $\protect\alpha <0$ and $\protect\nu \in
\lbrack 1,3]$. Right panel~: $\protect\alpha <0$ and increasing values of $%
\protect\nu >3$ from right to left. }
\end{figure}

The effective potential can thus have three different shapes depending on
the sign of $\alpha $, the value of the parameter $\nu $ and the location of
the horizon $x_{+}$~: either is is negative everywhere, or it is a negative
well edged by a barrier, at infinity if $\alpha >0$ or at the horizon if $%
\alpha <0$~; or, if $\alpha >0$, it may be a negative well surrounded by two
barriers, one near the horizon, the other at infinity. See figure~\ref%
{fig:Veff(u)} for two examples of such behaviours~; see figure~\ref%
{fig:VeffDecouplingLimit} of Appendix~\ref{app:DecouplingLimit} for a
typical shape of a negative effective potential.\newline

\begin{figure}[tbp]
\includegraphics[scale=0.4]{{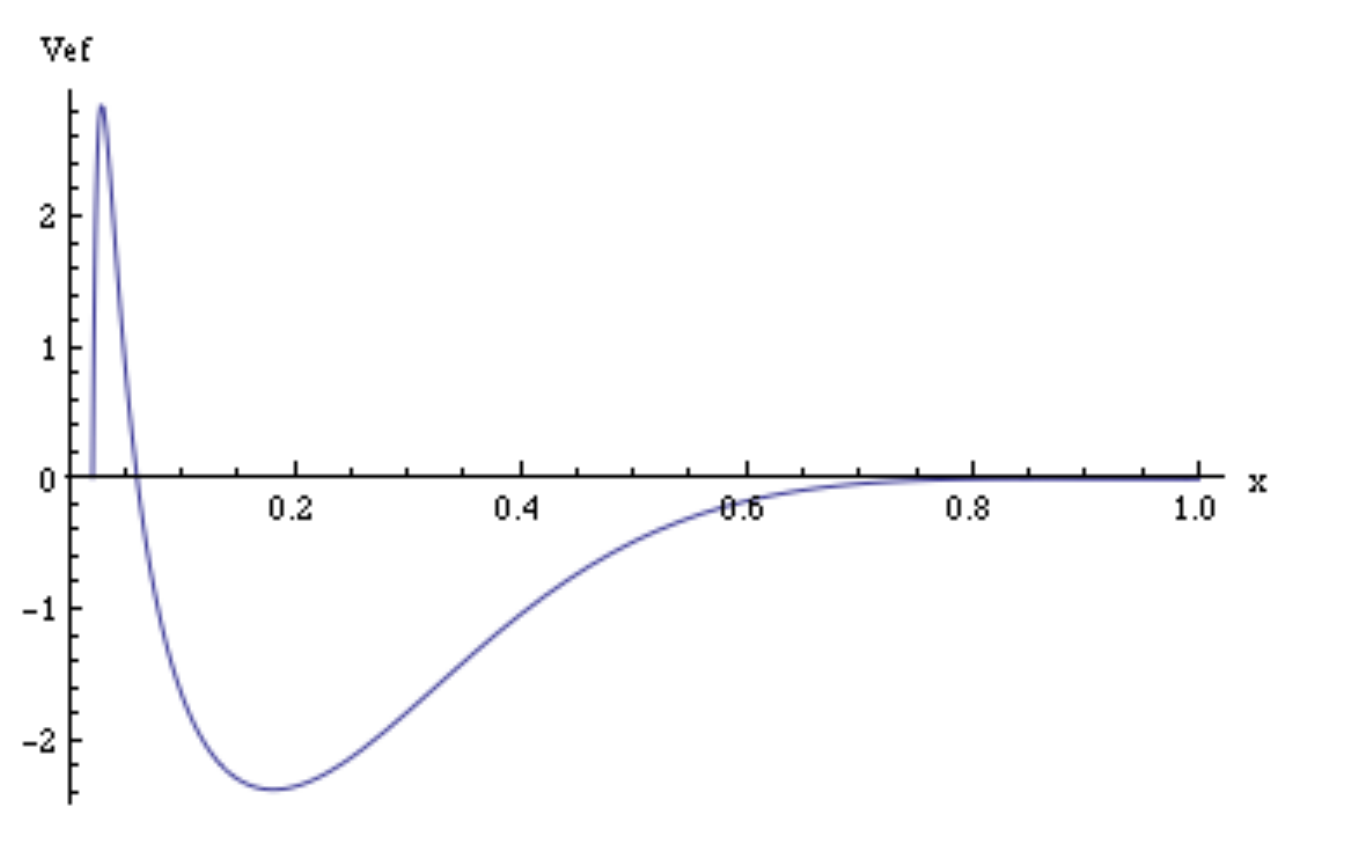}} %
\includegraphics[scale=0.4]{{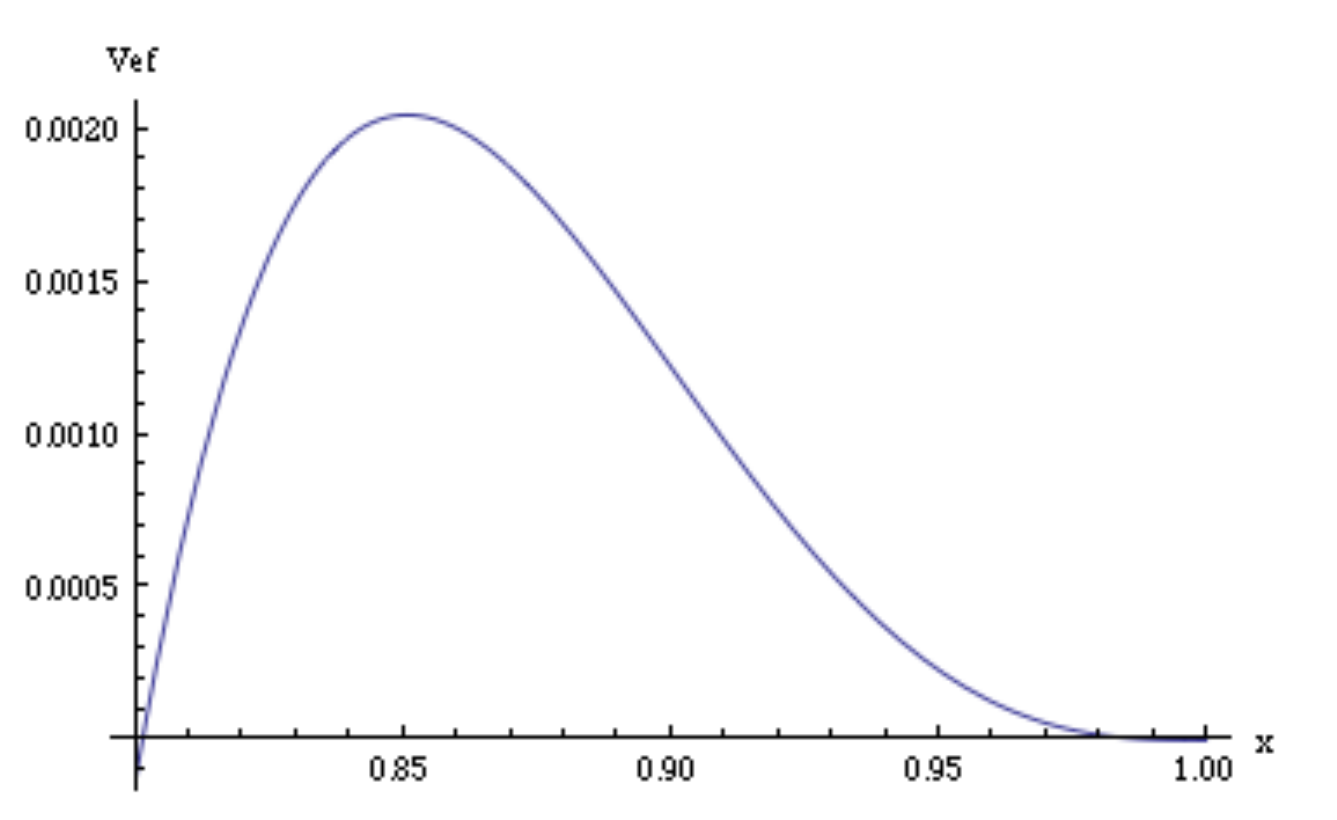}} %
\includegraphics[scale=0.4]{{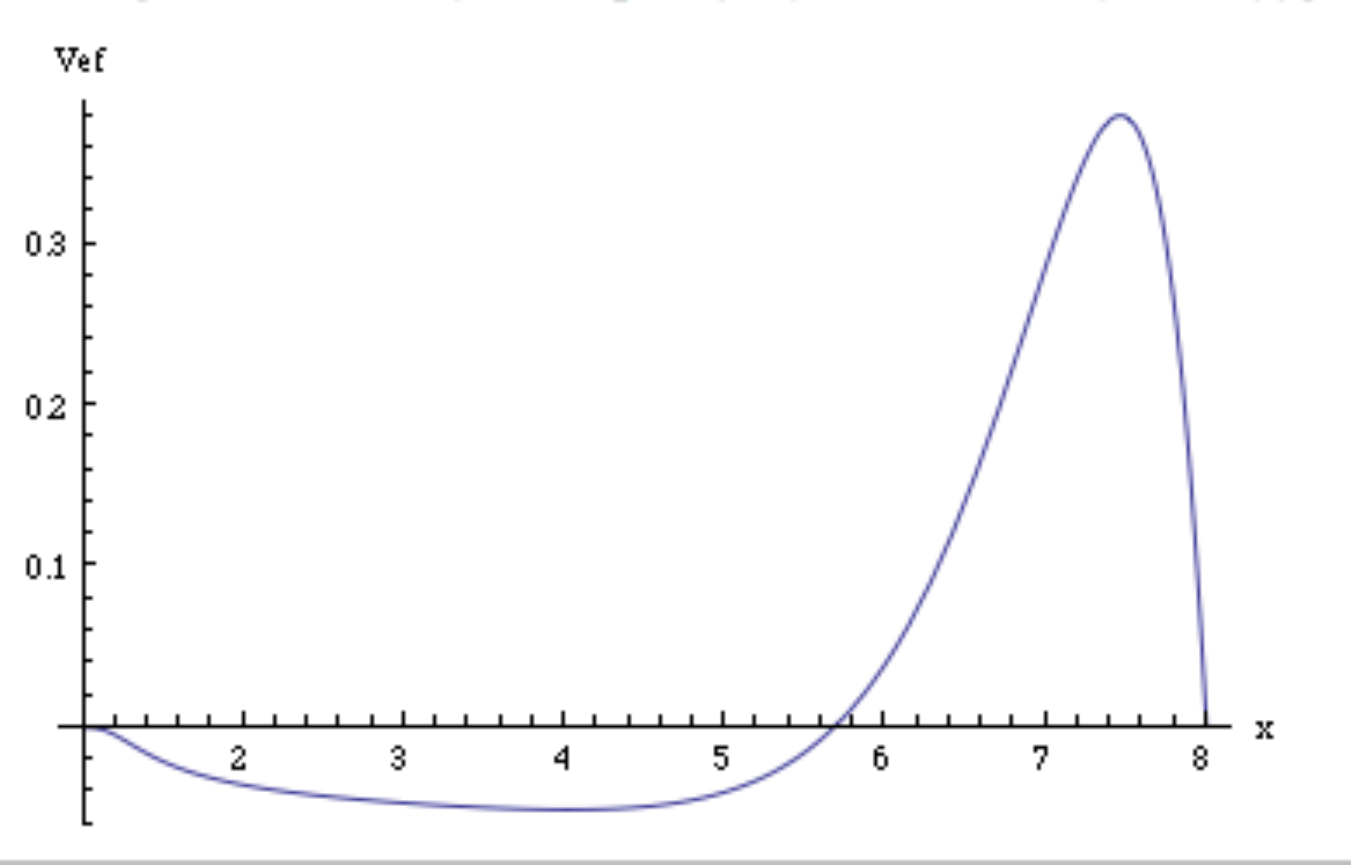}}
\caption{The effective potential $V_{\mathrm{eff}}$ as a function of the
radial coordinate $x$ (the explicit expression of which is given in Appendix~%
\protect\ref{app:Veff}). Left panel~: for $\protect\alpha>0$, $\protect\nu%
=1.95$ and $x_+=0.02$, $V_{\mathrm{eff}}$ is a negative potential well
surrounded by two barriers, one near the horizon $x_+$ and the other (zoomed
on the medium panel) at spatial infinity. Right panel~: for $\protect\alpha%
<0 $, $\protect\nu=8$ and $x_+=8$. }
\label{fig:Veff(u)}
\end{figure}

\subsection{The existence of bound states with negative $E^2$}

For values of $\nu $ and $x_{+}$ such that the effective potential is
everywhere negative then, as standard methods for solving the one
dimensional Shr\"{o}dinger equation show, the equation of motion (\ref{RP})
for the radial perturbations will possess bound states with \textsl{negative}
$E^{2}$. Since, for a negative value of $E^{2}$, the mode $\delta \phi $
blows up in time, see (\ref{FACT}), the black hole solution is then unstable.

\begin{figure}[tbp]
\includegraphics[scale=0.5]{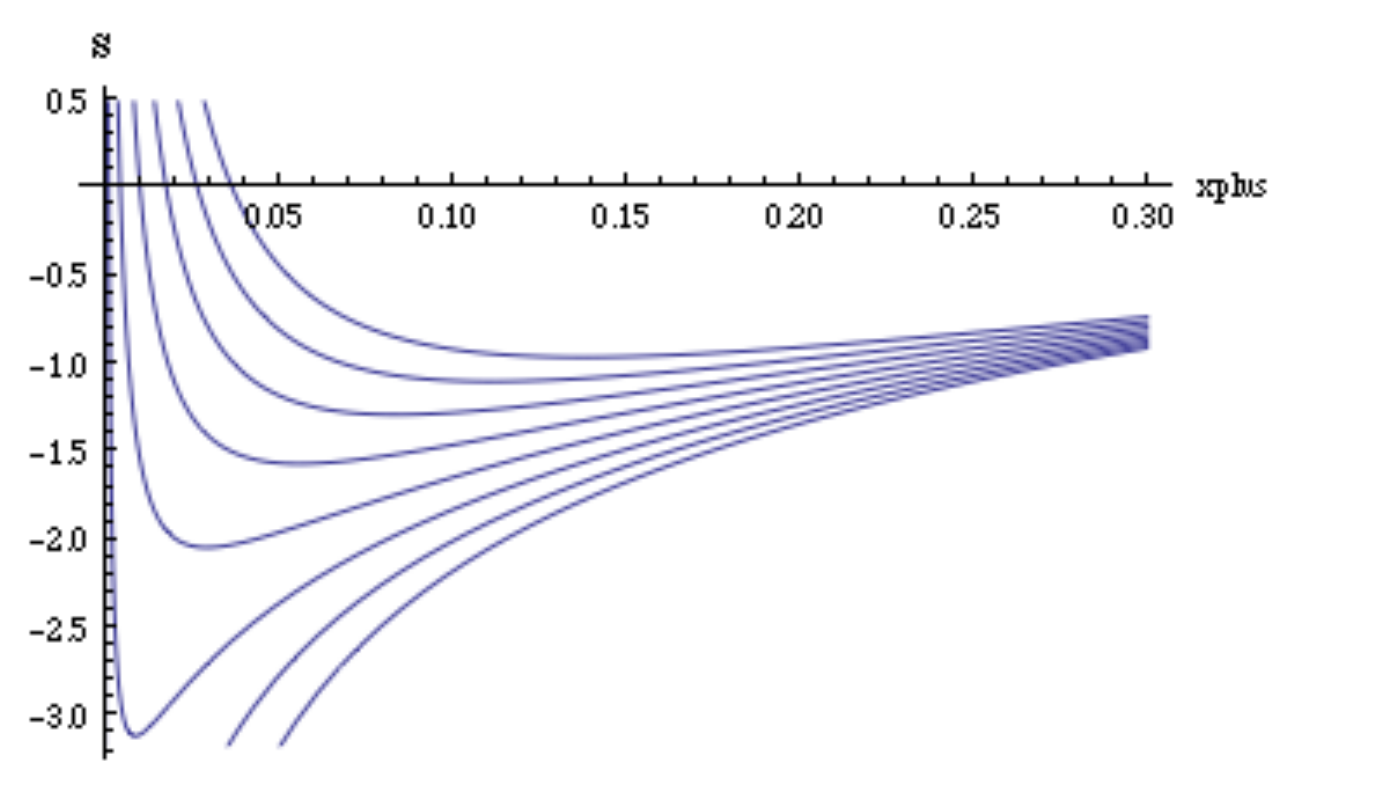} %
\includegraphics[scale=0.5]{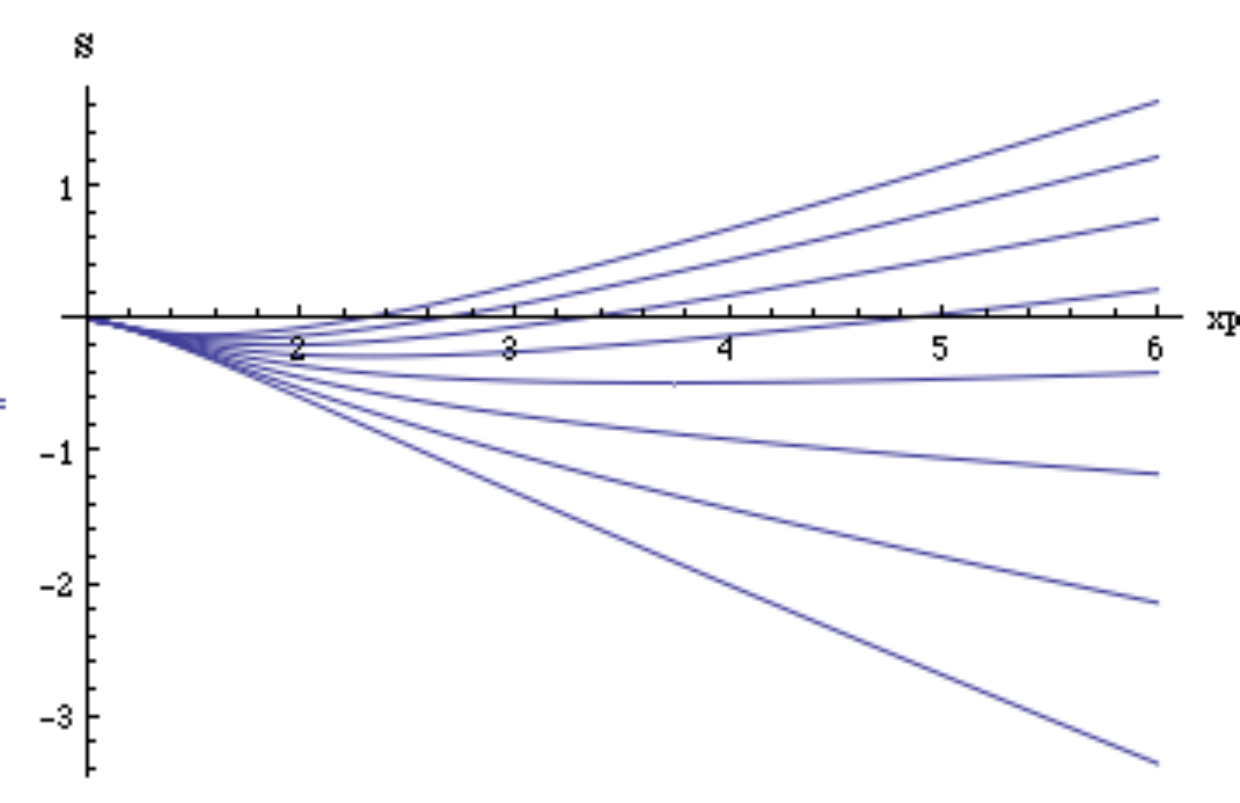}
\caption{Simon's integral $S$ as a function of the location of the horizon $%
x_{+}$ for various values of $\protect\nu $, for $\protect\alpha >0$ (left
panel) and $\protect\alpha <0$ (right panel). As can be seen it can be
positive, for all values of $\protect\nu $ if $x_{+}$ is small enough when $%
\protect\alpha >0$, and for $\protect\nu >7$ and $x_{+}$ big enough when $%
\protect\alpha >0$. }
\label{fig:SimonIntegral}
\end{figure}

For values of the parameters such that the effective potential is a negative
well surrounded by positive barriers near the horizon and/or at infinity,
then, in analogy with what happens in the case of a square potential well,
bounded by a potential barrier, there will be no bound state and the radial
perturbations will be stable, \textsl{if} the barrier is sufficiently high.
As shown by Simon~\cite{Simon1, Simon2} when $V_{\mathrm{eff}}$ is bounded
and fall off faster than $\left\vert \rho \right\vert ^{-2}$, a necessary
(but not sufficient) condition for the absence of bound states of negative $%
E^{2}$ is 
\begin{equation}
S\equiv \int_{-\infty }^{+\infty }V_{\mathrm{eff}}\,d\rho>0  \label{SIMON}
\end{equation}%
It turns out that $S$ can be given in closed form. Its explicit expression
is given in Appendix B. It is a function of $x_{+}$ and $\nu $ and its
behavior is shown on figure~\ref{fig:SimonIntegral}. Its asymptotic
behaviors are 
\begin{equation}
\begin{aligned} {S\over\sqrt{|\alpha|}}&\approx
-{\sqrt3\over2}(|1-x_+|)^{3\over2}\to0_-\qquad\hbox{for} \ x_+\to1_\pm\\
\alpha>0\ ,\ \nu\in[1,2]~:\quad {S\over\sqrt{\alpha}}&\approx
{\nu-1\over4\nu}\sqrt{2-\nu\over2+\nu}x_+^{-\nu/2}-{3\nu\over2\sqrt{4-%
\nu^2}}x_+^{\nu/2-1}\to+\infty\quad\hbox{ for}\  x_+\to0\\
\alpha>0,\nu>2~:\quad {S\over\sqrt\alpha}&\approx
{(\nu-1)\sqrt{\nu-2}\over4(\nu+2)x_+}\to+\infty\quad\hbox{ for}\  x_+\to0\\
\alpha<0~:\quad
{S\over\sqrt{|\alpha|}}&\approx-{7-\nu\over4\sqrt{2+\nu}}x_+\left(1+{\nu-1%
\over(7-\nu)x_+}\right)\to\pm\infty\quad\hbox{for}\ x_+\to+\infty
\end{aligned}
\end{equation}%
When $\alpha >0$, $S$ is positive for all values of $\nu $ for small enough
values of $x_{+}$. When $\alpha <0$, $S$ is positive for all values of $\nu
>7$ for large enough values of $x_{+}$. Hence there is a possibility that
there be no bound states in these cases and that the black hole solutions
then be stable.\newline

Concentrating on cases when the Simon integral (\ref{SIMON}) is positive we
hence integrated numerically the differential equation (\ref{RP}) using the
standard \textquotedblleft shooting" method (see \textsl{e.g.} \cite{Nath1}). It consists in choosing the boundary value for $u(\rho )$ and its
derivative in the potential barrier near the horizon in order to guarantee
that the solution exponentially decreases as one approaches the horizon. A
range of values for $E^{2}$ is then scanned (with $E^{2}$ negative but
smaller than the depth of the well). If the corresponding mode functions $%
u(\rho ,E^{2})$ \textsl{all} blow up exponentially for large $\rho $ to,
say, $+\infty $, then the equation has \textsl{no} bound state. If, on the
contrary, $u(\rho ,E_{1}^{2})$ goes to $+\infty $ but $u(\rho ,E_{2}^{2})$
goes to $-\infty $ then a bound state exists for a value $E_{n}^{2}\in
\lbrack E_{1}^{2},E_{2}^{2}]$.

We did not explore the whole range of possibilities
and limited ourselves to \textquotedblleft reasonable" values of $\nu $ and $%
x_{+}$. Within these cases we were not able to find values for the parameters $\nu $ and $x_{+}$
which would \textsl{not} exhibit any bound state, see figure~\ref%
{fig:boundstate} in Appendix~\ref{app:DecouplingLimit} for typical
behaviors of the mode functions $u$.\footnote{%
For example~: for $\alpha =1$, $\nu =1.7$, $x_{+}=0.001$, and hence $\eta
^{2}=17.3$ and $S=5$, there is a bound state for $E^{2}$ such that $%
2.1<|E^{2}|<2.2$. For $\alpha >0$, $\nu =6$, $x_{+}=0.05$, and hence $\eta
^{2}\alpha ^{-1}=0.25$ and $S=4.6$, there is a bound state for $E^{2}$ such
that $0.11<|E^{2}|<0.12$. Finally, for $\alpha =-1$, $\nu =9$, and hence $%
\eta ^{2}=0.09$ and $S=0.45$, there is a bound state for $E^{2}$ such that $%
0.110<|E^{2}|<0.115$.}

Therefore the class of hairy black holes (2.4) are \textsl{unstable}, in
that the spectrum of their spherically symmetric perturbations exhibits, at
least generically, modes which grow exponentially in time.

However, as we shall now see, this instability can be marginal for small
black holes.

\section{\label{sec:savethephenomena}A class of marginally stable small
black holes}

A first remark is, as was already established below eq. (\ref{BC2}), is that
the equation (\ref{RP}) for the radial perturbations possesses only a 
\textsl{finite} number of unstable bound states. Actually, our numerical exploration showed that for every value of the parameters there is only {\sl one} bound state.

A second remark is that the eigenvalues of these bound states are bounded by
a sharp Lieb-Thirring inequality \cite{Hundertmark:1998cz}%
\begin{equation}
\sum_{i}\sqrt{-E_{i}^{2}}\leq \frac{1}{2}\int_{V_{\mathrm{eff}}<0}\left\vert
V_{\mathrm{eff}}\right\vert d\rho ,  \label{I1}
\end{equation}%
where the sum is for all $E_{i}^{2}<0$. Now, as is pointed out in Appendix
A, $V_{\mathrm{eff}}$ is proportional to $\alpha $, therefore we can take $%
E^{2}=|\alpha |\bar{E}^{2}$ where $\bar{E}^{2}$ is a number, of order unity
if the dimensionless parameters $\nu ,$ $x_{+}$ and $\eta ^{2}\alpha ^{-1}$
are all taken of order unity (see examples of values of $\bar{E}^{2}$ in
footnote 2). In that case the unstable mode 
\begin{equation}
\delta \phi \propto e^{\mathrm{i}Et}=e^{+\sqrt{-\bar{E}^{2}}\,\sqrt{\alpha }%
\,t}
\end{equation}%
then grows on a time scale $t$ of the order of $1/\sqrt{\alpha }$. Since the
characteristic time scale of gravitational effects is set by the mass $m$ of
the black hole, the growth of the modes will be tamed if $\sqrt{\alpha }\,m$
is much less than 1. Returning to the relationship between the mass of the
black hole and its horizon size given in (2.9) we therefore see that this
condition is met if $\alpha >0$ and $\nu \in \lbrack 1,2]$ in which case the
mass and size of the black hole are related by 
\begin{equation}
2\nu \sqrt{\alpha }\,m\approx (4-\nu ^{2})^{(\nu -1)/2}(\sqrt{\alpha }%
\,r_{+})^{2-\nu }\ll 1\,.
\end{equation}%
Such black holes are therefore marginally stable, that is, \textquotedblleft
long-lived".

\section{Conclusions}

We studied in this paper the stability of the asymptotically flat black
holes with minimally coupled scalar hair discovered in~\cite{Andrespapers1,
Andrespapers2, Andrespapers3}. We limited ourselves to the analysis of the
radial \textquotedblleft s-wave" perturbations of the metric and the scalar
field, which can be presumed to be the least stable (since a stabilizing
barrier in $l(l+1)/r^{2}$ appears in the effective potential of $l\neq 0$
modes). We found that, generically, some of these radial modes were
spatially bounded but growing in time. Strictly speaking the black holes are
therefore unstable. However, if their mass $m$ and size $r_{+}$ are small
compared with the scale $1/\sqrt{\alpha }$ of the model the characteristic
time for the instability to develop is long compared to the scale set by $m$.

A final remark is in order. Since the potential for the scalar field is
unbounded from below, it is tempting to say that the instability could have
been foreseen without any detailed analysis. However this would have been a
hasty conclusion (as the well-known example of anti-de Sitter spacetime
already teaches us~\cite{AdSBH}). In fact, when a proper cosmological term
is added to the potential, leaving it unbounded from below (see~\cite%
{Andrespapers1, Andrespapers2, Andrespapers3} for its expression), so that
the black hole is asymptotically anti-de Sitter, it is easy to see that the
effective potential $V_{\mathrm{eff}}$ for the perturbations can be \textsl{%
positive} everywhere, so that the black hole is linearly stable against
radial perturbations. We leave a thorough study of this promising case to
further work.

While completing
this work, an article that analyzes a similar situation and reaches
conclusions which are in harmony with ours was published \cite%
{Kleihaus:2013tba}.

\begin{acknowledgments}
A.A. thanks Yves Brihaye for interesting discussions about gravitational perturbations and to Barry Simon for useful discussions about his work. Research of A.A. is supported in part by the FONDECYT grant 11121187.
N.D. acknowledges financial support from CONICYT and thanks the  Adolfo Iba$\tilde {\rm n}$ez University of Vi$\tilde{\rm n}$a del Mar and the Pontificia Universidad Cat\'olica de Valparaiso for their hospitality during her most enjoyable stay in Chile.

\end{acknowledgments}

\appendix

\section{\label{app:Veff}Explicit expression of the effective potential $V_{%
\mathrm{eff}}$}

The effective potential (3.6) using (3.7) and (2.4) is~: 
\begin{equation}
V_{\mathrm{eff}}=F\,{\frac{\mathrm{N_{1}}+x^{2}\mathrm{N_{2}}}{\mathrm{D}}}
\end{equation}%
where $F$ is the metric function given in (2.4), where 
\begin{equation}
\left\{ \begin{aligned} {\rm
D}&=4\beta\nu^2(4-\nu^2)\,x^{2+\nu}\,[(\nu-1)+(\nu+1)x^\nu]^2\,, \\ {\rm
N_1}&=c_1 x^2+c_2 x^\nu+c_3 x^{2\nu}+c_4x^{3\nu}\quad,\quad {\rm
N_2}=d_1x^\nu+d_2x^{2\nu}+d_3x^{3\nu}+d_4x^{4\nu} \end{aligned}\right.
\end{equation}%
and where the coefficients are given by 
\begin{equation}
\left\{ \begin{aligned} c_1&=-[1+\beta(2-\nu)](\nu+2)(\nu-1)^4\quad,\quad
c_2=(3-\nu)(\nu-1)^3\nu^2\\ c_3&=6\nu^2(\nu^2-1)^2\quad,\quad
c_4=-\nu^2(1+\nu)^3(3+\nu) \end{aligned}\right.
\end{equation}

\begin{equation}
\left\{ \begin{aligned}
d_1&=(\nu-1)(\nu+2)[-4+5\nu-17\nu^2-\nu^3+\nu^4-4\beta(2-\nu)(2\nu^3+2\nu^2-%
\nu+1)]\\ d_2&=6(\nu^2-1)[2+11\nu^2-\nu^4+\beta(4-\nu^2)(1+3\nu^2)] \\
d_3&=(\nu+1)(\nu-2)[-4-5\nu-17\nu^2+\nu^3+\nu^4-4\beta(2+\nu)(-2\nu^3+2%
\nu^2+\nu+1)]\\ d_4&=(\nu-2)(\nu+1)^4[1+\beta(2+\nu)]\,. \end{aligned}\right.
\end{equation}%
where $\eta ^{2}=\beta \alpha $. Finally $\beta $ can be writen in terms of $%
\nu $ and $x_{+}$ using the expresion given in (2.7). Hence $V_{\mathrm{eff}%
}/\alpha $ is a function of the radial coordinate $x$, the parameter $\nu $
and the location of the horizon $x_{+}$.

For $\nu =2$ (A.1-6) reduces to the effective potential studied in Appendix~%
\ref{app:nu=2}. For $\nu =1$ it reduces to the \textquotedblleft decoupling
limit" effective potential studied in Appendix~\ref{app:DecouplingLimit}.%
\newline

\section{\label{app:SimonIntegral}The Simon integral}

Using (3.9) and the explicit expression for $V_{\mathrm{eff}}$ given
Appendix A, we have that the Simon integral (3.13) reads 
\begin{equation}
S=\pm\int_{x_+}^1V_{\mathrm{eff}}{\frac{d\rho}{dx}}dx=\pm\int_{x_+}^1{\frac{%
\eta\,(\mathrm{N_1}+x^2\mathrm{N_2})}{\mathrm{D}}}dx=\pm{\frac{\alpha}{\eta}}%
\,{\frac{\mathrm{S_1}}{\mathrm{S_2}}}\,,
\end{equation}
where the plus sign holds for $\alpha>0$ and the minus sign for $\alpha<0$,
where $\beta$ is given in terms of $x_+$ and $\nu$ in (2.7) and where 
\begin{equation}
\begin{aligned} {\rm
S_1}=&(\nu-1)(2-\nu)+2(\nu+1)(\nu+2)x_+-(\nu+2)(\nu+7)x_+^2\\
&-2x_+^{\nu-1}[3\nu^2-2x_+(1+2\nu^2)-x_+^2(\nu^2-4)-2x_+^3(7-\nu^2)]\\
&-x_+^{2\nu}[(\nu+1)(\nu+2)+2(2-\nu)(\nu-1)x_++(7-\nu)(2-\nu)x_+^2]
\end{aligned}
\end{equation}
and 
\begin{equation}
\mathrm{S_2}=4(4-\nu^2)x_+(1-x_+^\nu)^2\,.
\end{equation}

\section{\label{app:nu=2} The case $\protect\nu=2$}

When $\nu=2$ the general potential (2.3) and metric functions (2.4) have the
following limits (with $\eta^2=\alpha\beta$)~: 
\begin{equation}
\begin{aligned}
V&={\alpha\over16}(-12\psi\cosh\psi+9\sinh\psi+\sinh3\psi)\qquad\hbox{with}%
\qquad\psi=\sqrt{2\over3}\,\phi\\ \Omega&={4x\over\alpha
\beta(1-x^2)^2}\qquad,\qquad F={\alpha\over16}\left[(1-x^2)(3-x^2)+4\ln
x+4\beta(1-x^2)^2\right]\,. \end{aligned}
\end{equation}
The horizon of the black hole $x_+$ is such that $F(x_+)=0$ and the
parameter $\beta$ can thus be traded for its expression in terms of $x_+$~: 
\begin{equation}
\beta=-{\frac{(x_+^2-1)(x_+^2-3)+4\ln x_+}{4(1-x_+^2)^2}}\,.
\end{equation}
The explicit expression of the effective potential $V_{\mathrm{eff}}$ (3.6)
is not very illuminating~: it is given by the limit of (A.1-5) for $\nu\to2$%
. For $\alpha<0$ it is everywhere negative. For $\alpha>0$ its behaviour for 
$x_+$ small (e.g. $x_+=0.02$) is quite similar to that of figure~\ref%
{fig:Veff(u)}. The Simon integral (3.13) is the limit of (B.1-3) for $%
\nu\to2 $ and reads 
\begin{equation}
\begin{aligned}
{S\over\sqrt\alpha}&={(1-x_+)\left[(1-x_+^2)(1+12x_++5x_+^2)+12x_+(2+x_+)\ln
x_+\right]\over8x_+(1+x_+)\sqrt{-4\ln x_+-(1-x_+^2)(3-x_+^2)}}\\
&\approx{1\over8x_+\sqrt{-3-4\ln x_+}}+{5+12\ln x_+\over4\sqrt{-3-4\ln
x_+}}\qquad\hbox{for}\quad x_+\to1\,. \end{aligned}
\end{equation}
$S$ has the same shape as in the generic case depicted in figure~\ref%
{fig:SimonIntegral} (apart from the fact that it goes to $+\infty$ when $%
x_+\to1$). It is positive for small enough $x_+$ (approximately, $x_+<0.01$).

The black holes are therefore unstable if $\alpha<0$. For $\alpha>0$ the
effective potential may exhibit a positive barrier but, as in the generic
case studied in the main text, we failed to find a range of $x_+$ for which
there is no bound state. The black holes are therefore unstable.

\section{\label{app:DecouplingLimit} The ``Decoupling Limit"}

Consider the Klein-Gordon equation~: 
\begin{equation}
\mathchoice{\vcenter{\vbox{\hrule height.4pt\hbox{\vrule
width.4pt height 5pt \kern 5pt\vrule width.4pt}\hrule height.4pt}}}{\vcenter{\vbox{\hrule height.4pt\hbox{\vrule
width.4pt height 5pt \kern 5pt\vrule width.4pt}\hrule height.4pt}}}{\vcenter{\vbox{\hrule height.3pt\hbox{\vrule
width.3pt height 2.1pt \kern 2.1pt\vrule width.3pt}\hrule height.3pt}}}{\vcenter{\vbox{\hrule height.3pt\hbox{\vrule
width.3pt height 1.5pt \kern 1.5pt\vrule width.3pt}\hrule height.3pt}}}\psi-{%
\frac{dV}{d\psi}}=0\quad\hbox{with}\quad V={\frac{2\alpha}{3}}%
[\sinh\psi(\cosh\psi-4)+3\psi]\quad\hbox{so that}\quad {\frac{dV}{d\psi}}={%
\frac{4\alpha}{3}}(\cosh\psi-1)^2\,.
\end{equation}
This is an odd potential which is very flat at $\psi=0$ ($V=\alpha\psi^5/15+%
\mathcal{O}(\psi^6)$) and unbounded from below, of the type shown on figure~%
\ref{fig:PotAndres}.

On a Schwarzschild background, 
\begin{equation}
ds^2=(1-2m/r)dt^2+{\frac{dr^2}{1-2m/r}}+r^2(d\theta^2+\sin^2\theta d\phi^2),
\end{equation}
(where $m$ is a positive mass parameter), the field 
\begin{equation}
\psi=\ln[1+1/(\eta r)],\qquad\hbox{with $\eta$ such that}\qquad
6\eta^3m+3\eta^2+\alpha=0,
\end{equation}
solves the above Klein-Gordon equation (C.1). For $\psi$ to be bounded
everywhere but at the singularity $r=0$, $\eta$ must be positive. Hence $%
\alpha<-3\eta^2$ must be negative.\newline

The potential (D.1) and the Schwarzschlid metric (D.2) are the limit of the
general potential and metric functions (2.3-4) when $\nu\to1$, with $%
x\equiv1+1/(\eta r)$.\newline

At linear order the Schwarzschild metric is left unperturbed and, setting $%
\psi=\ln[1+1/(\eta r)]+\delta\psi(t,r,\theta,\phi)$, the perturbation $%
\delta\psi$ satisfies the following linear equation~: 
\begin{equation}
\mathchoice{\vcenter{\vbox{\hrule height.4pt\hbox{\vrule
width.4pt height 5pt \kern 5pt\vrule width.4pt}\hrule height.4pt}}}{\vcenter{\vbox{\hrule height.4pt\hbox{\vrule
width.4pt height 5pt \kern 5pt\vrule width.4pt}\hrule height.4pt}}}{\vcenter{\vbox{\hrule height.3pt\hbox{\vrule
width.3pt height 2.1pt \kern 2.1pt\vrule width.3pt}\hrule height.3pt}}}{\vcenter{\vbox{\hrule height.3pt\hbox{\vrule
width.3pt height 1.5pt \kern 1.5pt\vrule width.3pt}\hrule height.3pt}}}%
\delta\psi-{\frac{d^2V}{d\psi^2}}\Big\vert_b\,\delta\psi=0\,,\qquad%
\hbox{where}\qquad{\frac{d^2V}{d\psi^2}}\Big\vert_b=-{\frac{2(1+2\eta
m)(1+2\eta r)}{r^2(1+\eta r)^2}}
\end{equation}
is $d^2V/d\psi^2$ evaluated on the background solution (D.3), where we have
traded $\alpha$ for $\alpha=-3\eta^2(1+2\eta m)$, and where $%
\mathchoice{\vcenter{\vbox{\hrule height.4pt\hbox{\vrule
width.4pt height 5pt \kern 5pt\vrule width.4pt}\hrule height.4pt}}}{\vcenter{\vbox{\hrule height.4pt\hbox{\vrule
width.4pt height 5pt \kern 5pt\vrule width.4pt}\hrule height.4pt}}}{\vcenter{\vbox{\hrule height.3pt\hbox{\vrule
width.3pt height 2.1pt \kern 2.1pt\vrule width.3pt}\hrule height.3pt}}}{\vcenter{\vbox{\hrule height.3pt\hbox{\vrule
width.3pt height 1.5pt \kern 1.5pt\vrule width.3pt}\hrule height.3pt}}}$ is
the d'Alembertian on the Schwarzschild metric.

\medskip We decompose $\delta\psi$ in Fourier modes and spherical harmonics,
and introduce the tortoise coordinate: 
\begin{equation}
\delta\psi=e^{-\mathrm{i}Et}\,Y^l_m(\theta,\phi){\frac{u(r)}{r}}%
\qquad,\qquad \rho=r+2m\ln({r/2m-1})
\end{equation}
so that the perturbed Klein-Gordon equation (D.4) reduces to 
\begin{equation}
{\frac{d^2u}{d\rho^2}}=(V_{\mathrm{eff}}-E^2)u\qquad\hbox{with}\qquad V_{%
\mathrm{eff}}=\left(1-{\frac{2m}{r}}\right)\left({\frac{d^2V}{d\psi^2}}%
\Big\vert_b+{\frac{l(l+1)}{r^2}}+{\frac{2m}{r^3}}\right)
\end{equation}
where $l$ is null or a positive integer and where $(d^2V/d\psi^2)|_b$ is
given in (D.4).

\begin{figure}[tbp]
\includegraphics[scale=0.6]{{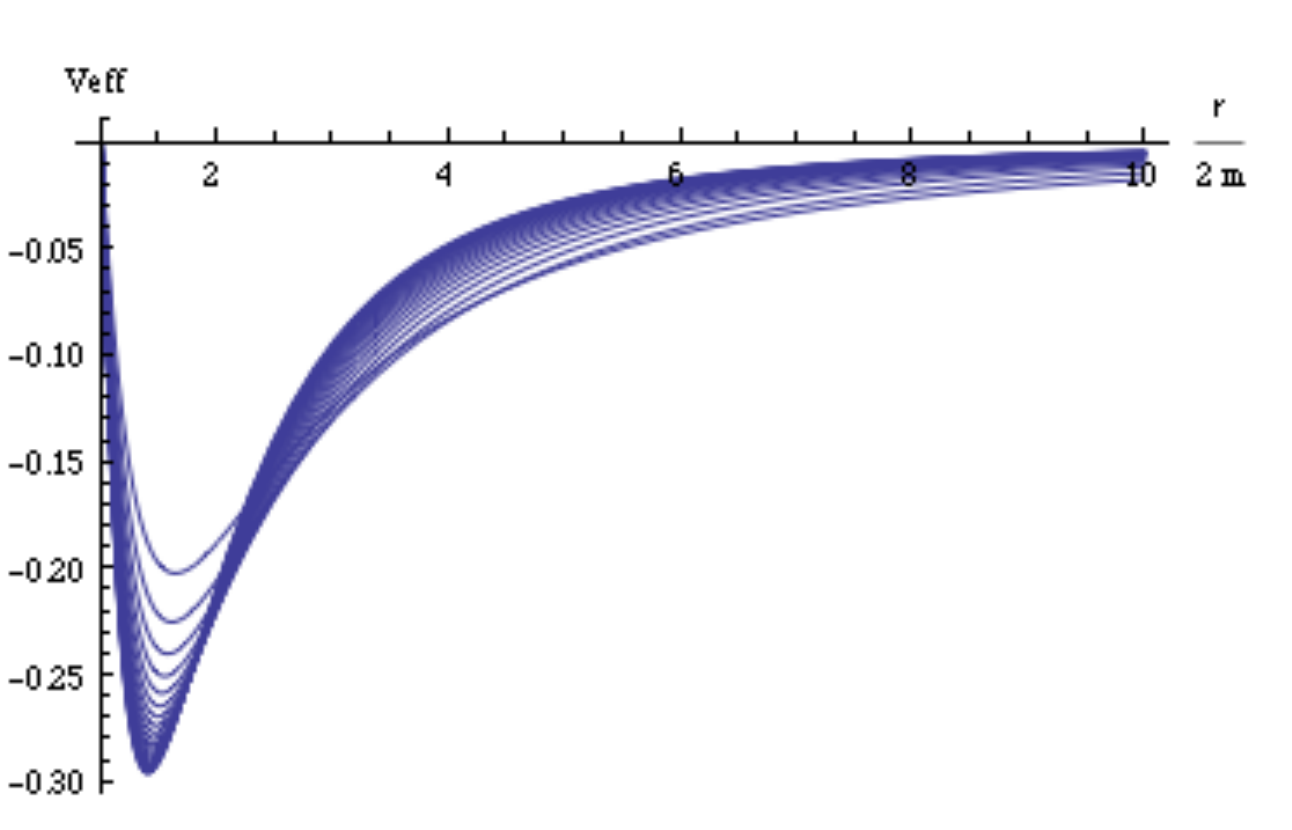}} %
\includegraphics[scale=0.6]{{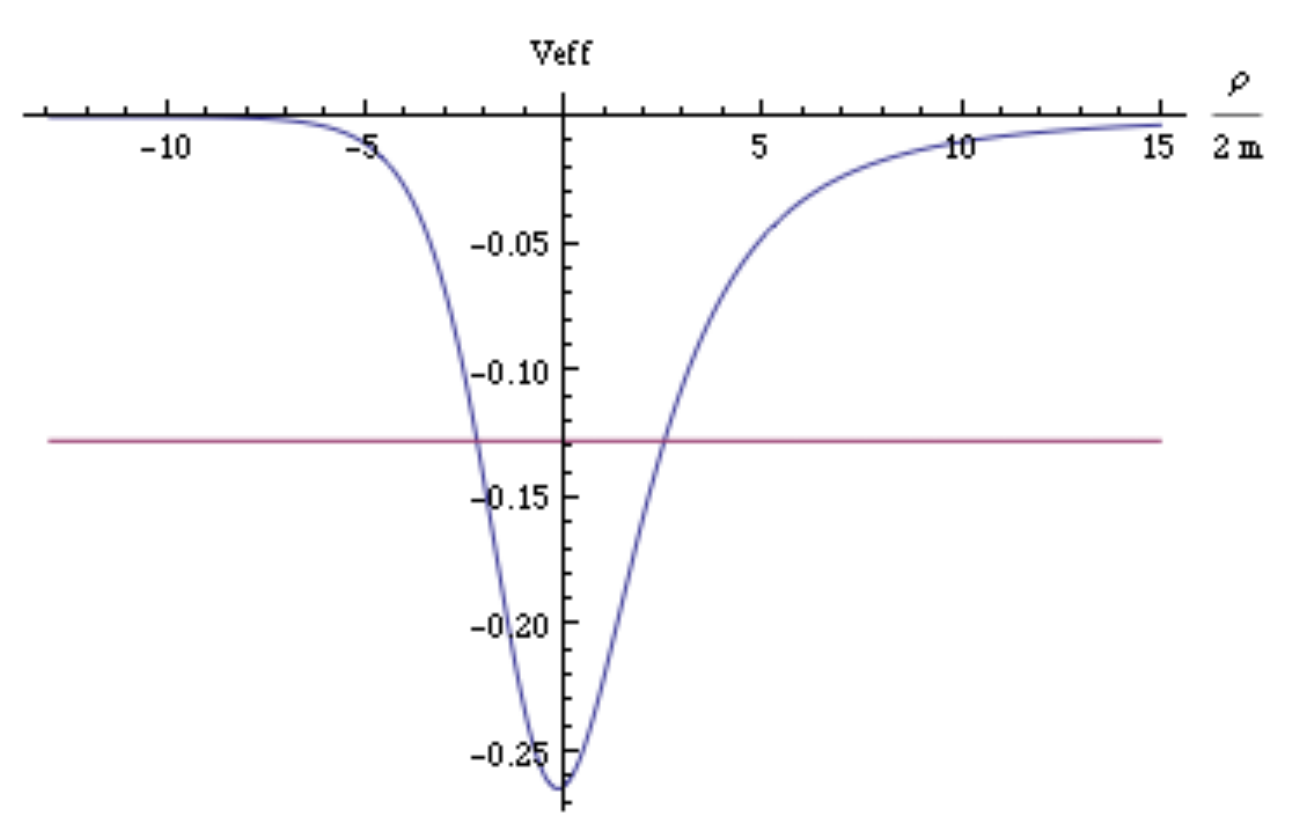}}
\caption{The ``decoupling limit" effective potential $V_{\mathrm{eff}}$ when 
$l=0$. Left panel~: for various values of the parameter $\protect\eta$ and
in fonction of $r$~; rightpanel~: for $\protect\eta=0.5$ and in function of
the tortoise coordinate $\protect\rho$. Also is shown the (negative) value
of $E^2$ corresponding to the bound state of figure~\protect\ref%
{fig:boundstate}. }
\label{fig:VeffDecouplingLimit}
\end{figure}

When $l=0$, $V_{\mathrm{eff}}$ is \textsl{negative} for all $r>2m$ (for all
positive $m$ and positive $\eta$), see figure~\ref{fig:VeffDecouplingLimit}.
Bound states with \textsl{negative} $E^2$ therefore exist, which is
confirmed by a numerical analysis of the solutions of (D.6), see figure~\ref%
{fig:boundstate}.

\begin{figure}[tbp]
\includegraphics[scale=0.6]{{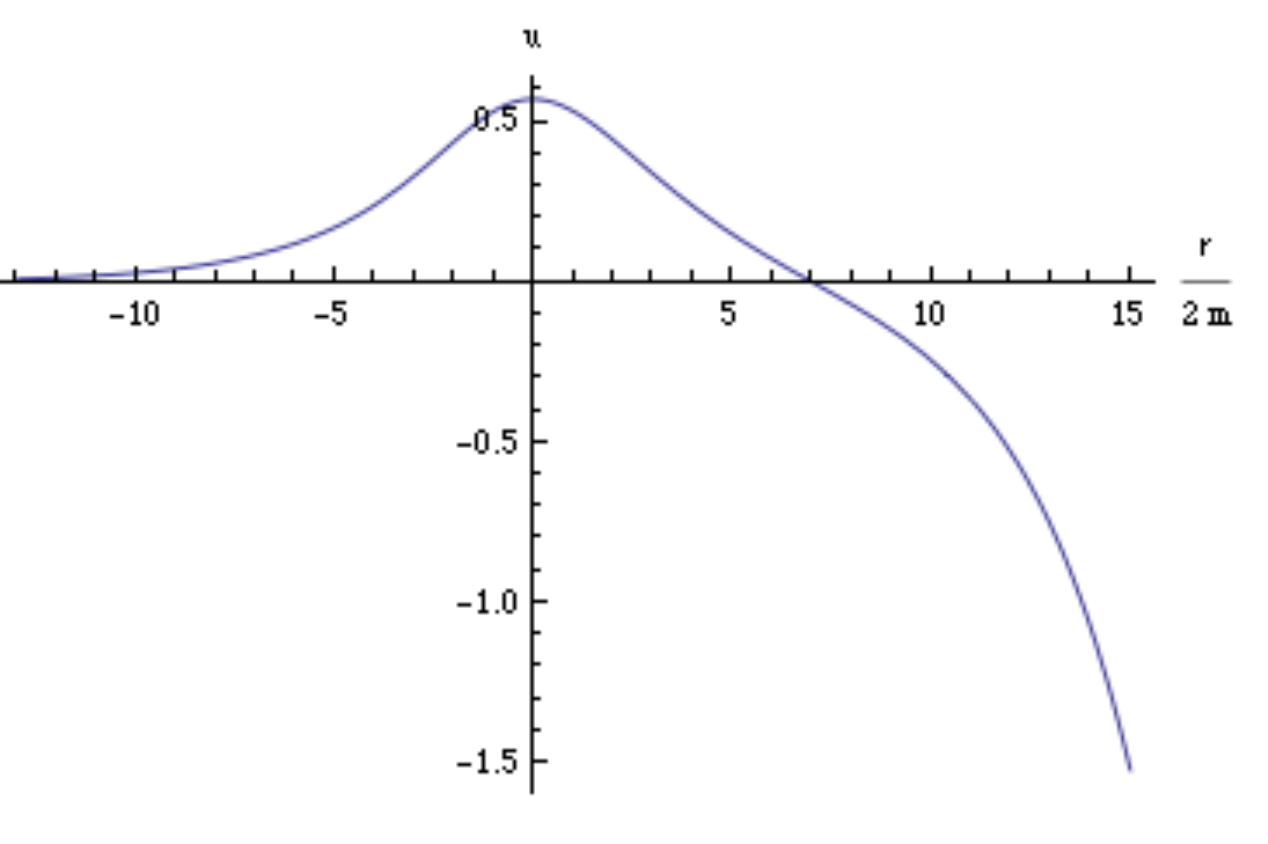}} %
\includegraphics[scale=0.6]{{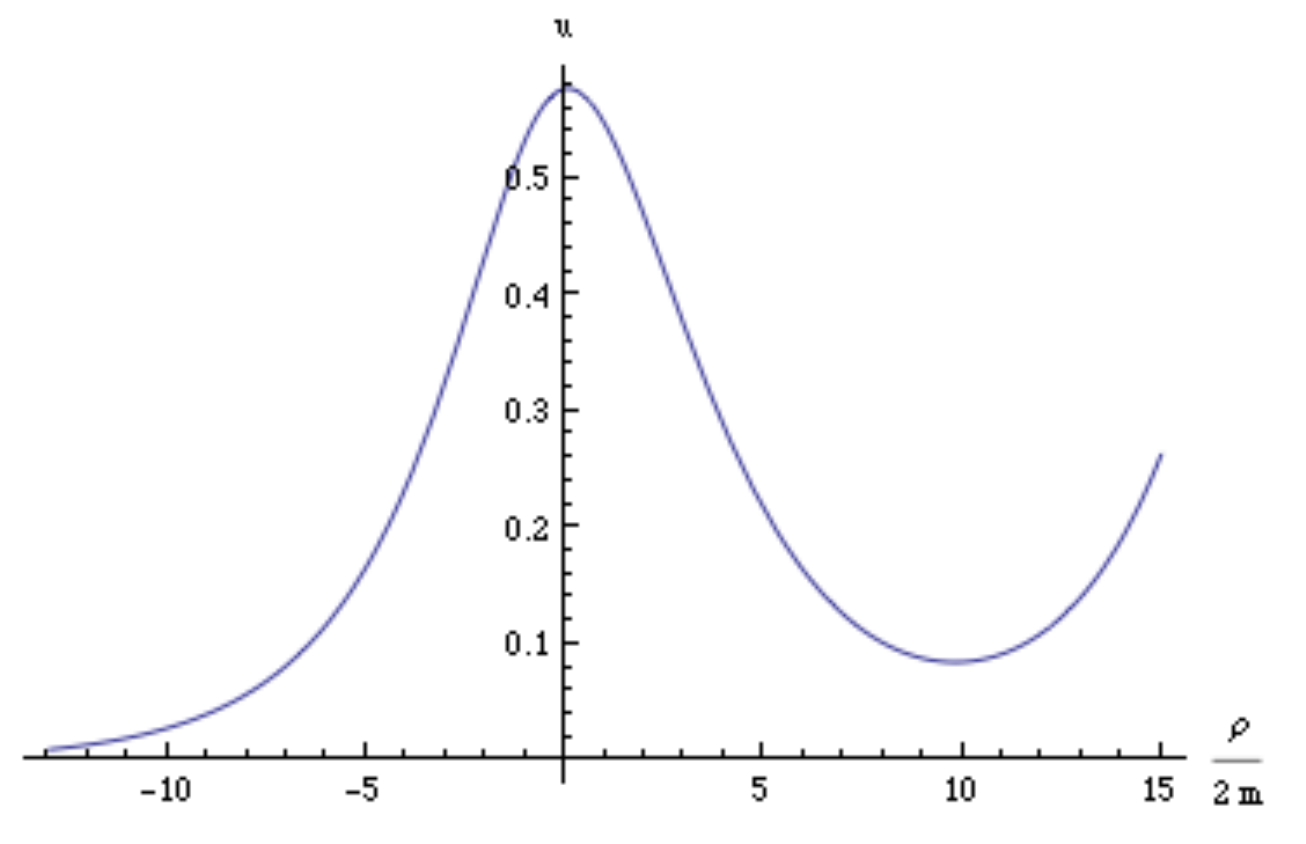}}
\caption{Solution of Eq. (D.6) for $\protect\eta=0.5$. Left panel~: for $%
E_+^2=-0.122$~; right panel~: for $E_-^2=-0.127$. A solution, bounded at the
horizon and at infinity, exists for an eigenvalue $E_1^2$ such that $%
E_-^2<E_1^2<E_+^2<0$. }
\label{fig:boundstate}
\end{figure}

\bigskip For a negative value of $E^2$, the mode $\delta\psi$ blows up in
time. Therefore the solution (D.3) of the Klein-Gordon equation (D.1) is
linearly \textsl{unstable} on the Schwarzschild background (D.2), if one
imposes $m>0$ and that the background solution (D.3) diverges only at $r=0$,
which requires $\eta>0$.

\end{document}